\documentclass[11pt]{article}
\usepackage{amsmath}
\usepackage{graphicx}
\date{}

\def\p{Painlev\'e}

\newcommand{\Bbb}[1]{\mbox{\msbm #1}\,}
\def\real{\Bbb R}

\font\msbm=msbm10

\newcommand{\beq}{\begin{equation}}
\newcommand{\eeq}{\end{equation}}
 
\medskip

\def\d{\delta}

\def\z{\zeta}

\def\p{\pi}

\def\i0{\int_0^{\infty}}

\def\p1{\frac{\partial}{\partial x_1}}
\def\p2{\frac{\partial}{\partial x_2}}
\def\p3{\frac{\partial}{\partial x_3}}

\def\d3{\frac{d}{d x_3}}
\def\q1{\frac{\partial}{\partial x_1}}
\def\q2{\frac{\partial}{\partial x_2}}
\def\z1{\frac{\partial}{\partial x_1}}

\begin{document}

\title{ Riemann-Hilbert Approach to the Helmholtz Equation in a quarter-plane. Revisited} 

\author{Alexander Its, Elizabeth Its \\
{\it Department of Mathematical Sciences},\\ 
{\it Indiana University -- Purdue University  Indianapolis}\\
{\it Indianapolis, IN 46202-3216, USA}}       

\maketitle

\vskip 0.5in 
\centerline{{\bf Abstract}}

We revisit the Helmholts equation in a quarter-plane in the framework 
of the Riemann-Hilbert approach to linear boundary value problems 
suggested in late  90s by A. Fokas. We show the role of
the Sommerfeld radiation condition in Fokas's scheme.

\section{Introduction}

This paper is a complement to our previous paper \cite{AIEIJK} as well as the
second author's paper \cite{g57}, where, following the general ideas of Fokas' method
\cite{g3}-\cite{g5}, we started to develop the Riemann-Hilbert scheme for solving the elastodynamic
equation in the quarter-plane. In \cite{AIEIJK}, we show that the problem can be reduced to the solution
of a certain matrix Riemann-Hilbert problem with a shift posed on a torus. A detail 
analysis of this problem is our ultimate goal. The modest objective of this paper is
to reveal the Riemann-Hilbert interpretation   of the Sommerfeld radiation condition in Fokas' scheme by
considering the more simple case of the Helmholts equation in a quarter-plane. 
 
\section{RH approach for Helmholts equation in a quarter-space}
The classical  boundary value problem for the Helmholtz equation in the quarter-plane
$(x,z), x\geq 0, z\geq 0$ is given as follows:

\beq\label{mh1}
u_{xx}+u_{zz}+h^2u=0,\quad u_x(z,0)=u_1, \quad u_z(0,x)=u_2
\eeq

 It's Lax pair  in terms of the
spectral parameter $\zeta$ has the following form:
\beq\label{mh2}
\phi_{z}-\frac{ih}{2}(\zeta +\frac{1}{\zeta})\phi=Q
\eeq
\beq\label{mh3}
\phi_{x}+\frac{h}{2}(\zeta -\frac{1}{\zeta})\phi={\tilde Q}
\eeq

where
\beq\label{mh4}
Q=\frac{1}{2}\tau-\frac{ih}{2\zeta}u,\;
{\tilde{Q}}=\frac{i}{2}\tau -\frac{h}{2\zeta}u
\eeq
where $\tau=u_z-iu_x$.
The spectral function $\phi$ is  limited and decaying as
\beq\label{mh5}
\phi=u+ O(\zeta)\quad\mbox{as}\; \zeta\rightarrow 0
\eeq
and
\beq\label{mh6}
\phi=\frac{i\tau}{h\zeta} \; \mbox{as} \;\zeta \rightarrow \infty
\eeq

Integrating along the three rays which are discussed in details in \cite{g3}- \cite{g5},  one obtains three solutions:

\beq\label{uif140}
\phi _1(\zeta,x,z) = \int_\infty^z e^{\frac{ih}{2}(\zeta
+\frac{1}{\zeta})(z-z')} Q(\zeta,z',x) dz'
\eeq
\beq\label{uif240}
\phi _2(\zeta,x,z) = \int_0^x e^{\frac{ih}{2}(\zeta
+\frac{1}{\zeta})z+\frac{h}{2}(\zeta -\frac{1}{\zeta})(x'-x)} {\tilde
Q}(\zeta,0,x') dx'+\int_0^z e^{\frac{ih}{2}(\zeta
+\frac{1}{\zeta})(z-z')} Q(\zeta,z',x) dz'
\eeq
 \beq\label{uif040}
\phi _3(\zeta,x,z) = \int_\infty^x e^{
\frac{h}{2}(\zeta -\frac{1}{\zeta})(x'-x)} {\tilde Q}(\zeta,z,x')
dx'  
\eeq
 and the jump functions
$\rho_{ij}, \,i,j =1,2,3$ 
\beq\label{uif401}
\rho_{13}(\zeta)=-\int_0^\infty e^{-\frac{ih}{2}(\zeta
+\frac{1}{\zeta})z'} Q(\zeta,z',0) dz'
+ \int_0^\infty e^{
\frac{h}{2}(\zeta -\frac{1}{\zeta})x'} {\tilde Q}(\zeta,0,x')
dx',\quad \rho_{31}=-\rho_{13}
\eeq
\beq\label{uif402}
\rho_{21}(\zeta)=-\phi_1(0,0)= \int_0^\infty e^{-\frac{ih}{2}(\zeta
+\frac{1}{\zeta})z'} Q(\zeta,z',0) dz'
\eeq
\beq\label{uif403}
\rho_{32}(\zeta)=\phi_3(\zeta,0,0)= - \int_0^\infty e^{
\frac{h}{2}(\zeta -\frac{1}{\zeta})x'} {\tilde Q}(\zeta,0,x')dx'
\eeq  
 As a result one can express the
spectral function $\phi$ as the following  Cauchy integral over the oriented
contour K presented on Figure  {\ref{f1}}:
\begin{figure}[ht]
\begin{center}
\resizebox{4in}{!}{ 
\includegraphics{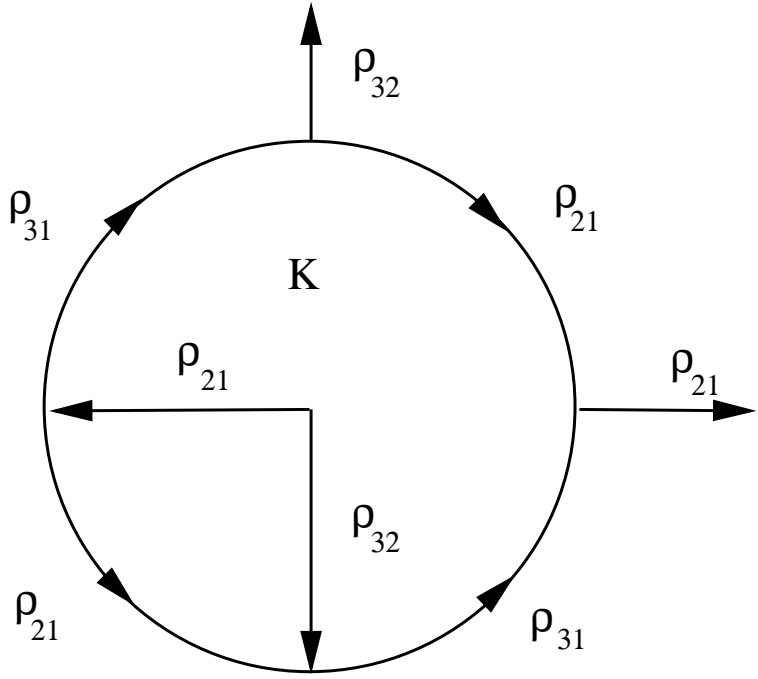}
}
\caption{Oriented contour $K$ for H equation}
\label{f1}
\end{center}
\end{figure}

\beq\label{h42a} 
\phi(x,z, \zeta)=\frac{1}{2\pi i}\int_1^\infty \frac{e^{\frac{ih}{2}(\zeta '
+\frac{1}{\zeta'})z-
\frac{ h}{2}(\zeta'-\frac{1}{\zeta'})x}\rho_{21}({\zeta'})d \zeta'}{\zeta'-\zeta}+
\frac{1}{2\pi i}\int_i^{i\infty} \frac{e^{\frac{ih}{2}(\zeta '
+\frac{1}{\zeta'})z-
\frac{ h}{2}(\zeta'-\frac{1}{\zeta'})x}
\rho_{32}({\zeta'})d \zeta'}{\zeta'-\zeta}
\eeq
$$
+\frac{1}{2\pi i}\int_0^{-i} \frac{e^{\frac{ih}{2}(\zeta '
+\frac{1}{\zeta'})z-
\frac{ h}{2}(\zeta'-\frac{1}{\zeta'})x}
\rho_{32}({\zeta'})d \zeta'}{\zeta'-\zeta}+
\frac{1}{2\pi i}\int_0^{-1} \frac{e^{\frac{ih}{2}(\zeta '
+\frac{1}{\zeta'})z-
\frac{ h}{2}(\zeta'-\frac{1}{\zeta'})x}
\rho_{21}({\zeta'})d \zeta'}{\zeta'-\zeta}
$$
$$+\frac{1}{2\pi i}\int_{C_1}\frac{e^{\frac{ih}{2}(\zeta '
+\frac{1}{\zeta'})z-
\frac{ h}{2}(\zeta'-\frac{1}{\zeta'})x}
\rho_{32}({\zeta'})d \zeta'}{\zeta'-\zeta}+
\frac{1}{2\pi i}\int_{C_2}\frac{e^{\frac{ih}{2}(\zeta '
+\frac{1}{\zeta'})z-
\frac{ h}{2}(\zeta'-\frac{1}{\zeta'})x}
\rho_{31}({\zeta'})d \zeta'}{\zeta'-\zeta}+
$$
$$
+\frac{1}{2\pi i}\int_{C_3} \frac{e^{\frac{ih}{2}(\zeta '
+\frac{1}{\zeta'})z-
\frac{ h}{2}(\zeta'-\frac{1}{\zeta'})x}
\rho_{32}({\zeta'})d \zeta'}{\zeta'-\zeta}+
\frac{1}{2\pi i}\int_{C_4} \frac{e^{\frac{ih}{2}(\zeta '
+\frac{1}{\zeta'})z-
\frac{ h}{2}(\zeta'-\frac{1}{\zeta'})x}
\rho_{31}({\zeta'})d \zeta'}{\zeta'-\zeta}
$$
where $C_1 \;-\; C_4$ are pieces of the circular part of $K$ in the first,second,
third and forth quadrants respectfully. Or, taking  ({\ref{mh5})  into account ( and changing
from $\zeta'$ to $\zeta$)

\beq\label{h421a} 
u(x,z, \zeta)=\frac{1}{2\pi i}\int_1^\infty \frac{e^{\frac{ih}{2}(\zeta 
+\frac{1}{\zeta})z-
\frac{ h}{2}(\zeta-\frac{1}{\zeta})x}\rho_{21}({\zeta})d \zeta}{\zeta}+
\frac{1}{2\pi i}\int_i^{i\infty} \frac{e^{\frac{ih}{2}(\zeta 
+\frac{1}{\zeta})z-
\frac{ h}{2}(\zeta-\frac{1}{\zeta})x}
\rho_{32}({\zeta})d \zeta}{\zeta}
\eeq
$$
+\frac{1}{2\pi i}\int_0^{-i} \frac{e^{\frac{ih}{2}(\zeta 
+\frac{1}{\zeta})z-
\frac{ h}{2}(\zeta-\frac{1}{\zeta})x}
\rho_{32}({\zeta})d \zeta}{\zeta}+
\frac{1}{2\pi i}\int_0^{-1} \frac{e^{\frac{ih}{2}(\zeta 
+\frac{1}{\zeta})z-
\frac{ h}{2}(\zeta-\frac{1}{\zeta})x}
\rho_{21}({\zeta})d \zeta}{\zeta}
$$
$$+\frac{1}{2\pi i}\int_{C_1}\frac{e^{\frac{ih}{2}(\zeta 
+\frac{1}{\zeta})z-
\frac{ h}{2}(\zeta-\frac{1}{\zeta})x}
\rho_{32}({\zeta})d \zeta}{\zeta}+
\frac{1}{2\pi i}\int_{C_2}\frac{e^{\frac{ih}{2}(\zeta 
+\frac{1}{\zeta})z-
\frac{ h}{2}(\zeta-\frac{1}{\zeta})x}
\rho_{31}({\zeta})d \zeta}{\zeta}
$$
$$
+\frac{1}{2\pi i}\int_{C_3} \frac{e^{\frac{ih}{2}(\zeta 
+\frac{1}{\zeta})z-
\frac{ h}{2}(\zeta-\frac{1}{\zeta})x}
\rho_{32}({\zeta})d \zeta}{\zeta}+
\frac{1}{2\pi i}\int_{C_4} \frac{e^{\frac{ih}{2}(\zeta 
+\frac{1}{\zeta})z-
\frac{ h}{2}(\zeta-\frac{1}{\zeta})x}
\rho_{31}({\zeta})d \zeta}{\zeta}
$$

Substituting the boundary conditions (\ref{mh1}})
into the jump functions (\ref{uif401}-\ref{uif403}) we obtain
\beq\label{mh61}
\rho_{31}=\frac{ih}{4}(\zeta-\frac{1}{\zeta})\int_0^\infty e^{\frac{-ih}{2}(\zeta+\frac{1}{\zeta})z}u(z,0)dz-
\frac{i}{2}\int_0^\infty e^{\frac{-ih}{2}(\zeta+\frac{1}{\zeta})z}u_1(z)dz
\eeq
$$\frac{h}{4}(\zeta+\frac{1}{\zeta})\int_0^\infty e^{\frac{h}{2}(\zeta-\frac{1}{\zeta})x}u(0,x)dx-
\frac{i}{2}\int_0^\infty e^{\frac{h}{2}(\zeta-\frac{1}{\zeta})x}u_2(x)dx$$
\beq\label{mh62}
\rho_{21}=\frac{ih}{4}(\zeta-\frac{1}{\zeta})\int_0^\infty e^{\frac{-ih}{2}(\zeta+\frac{1}{\zeta})z}u(z,0)dz-
\frac{i}{2}\int_0^\infty e^{\frac{-ih}{2}(\zeta+\frac{1}{\zeta})z}u_1(z)dz-\frac{1}{2}u(0,0)
\eeq
\beq\label{mh63}
\rho_{32}=\frac{h}{4}(\zeta+\frac{1}{\zeta})\int_0^\infty e^{\frac{h}{2}(\zeta-\frac{1}{\zeta})x}u(0,x)dx-
\frac{i}{2}\int_0^\infty e^{\frac{h}{2}(\zeta-\frac{1}{\zeta})x}u_2(x)dz+\frac{1}{2}u(0,0)
\eeq
The jump functions are not completely defined by the
known $u_1$ and $u_2$ functions, so one has to use
the global relationship
\beq\label{uif44} 
\rho_{21} + \rho_{32} =\rho_{31} \equiv 0.
\eeq
 to find the integrals of the unknown $u(z,0)$ and $u(0,x)$. Keeping
 for these integrals the same notations $\Phi_1$
and $\Phi_3$ as in \cite{AIEIJK}  we can write 
for
$$
|\zeta| \leq 1, \,\, 0\leq\arg\zeta\leq\frac{\pi}{2}, \quad \mbox{and}\quad
|\zeta| \geq 1, \,\, \pi\leq\arg\zeta\leq\frac{3\pi}{2}.
$$
 the following
global relationship
\beq\label{mh7}
\frac{ih}{4}(\zeta-\frac{1}{\zeta})\Phi_1(\zeta)-\frac{h}{4}(\zeta+\frac{1}{\zeta})
\Phi_3(\zeta)=F(\zeta)
\eeq
 where 
\beq\label{mh8}
\Phi_1(\zeta)=\int_0^{\infty}e^{\frac{-ih}{2}(\zeta+\frac{1}{\zeta})z}
u(0,z)dz
\eeq
\beq\label{mh9}
\Phi_3(\zeta)=\int_0^{\infty}e^{\frac{h}{2}(\zeta-\frac{1}{\zeta})x}
u(x,0)dx
\eeq
and
\beq\label{mh10}
F(\zeta)=\frac{i}{2}\int_0^{\infty}e^{\frac{h}{2}(\zeta-\frac{1}{\zeta})x}
u_2(x)dx+\frac{i}{2}\int_{\infty}^0 e^{-\frac{ih}{2}(\zeta +\frac{1}{\zeta})z}
u_1(z)dz
\eeq

Using the symmetry on can extend ({\ref{mh7}}) to the  complex plane $\zeta$
as follows:
\beq\label{mh12}
\zeta\,\rightarrow\,-\frac{1}{\zeta}:\;
\frac{ih}{4}(\zeta-\frac{1}{\zeta})\Phi_1(-\zeta)+
\frac{h}{4}(\zeta+\frac{1}{\zeta})
\Phi_3(\zeta)={ F(-{\frac{1}{\zeta}})}
\eeq
\beq\label{mh13}
\zeta\,\rightarrow\,\frac{1}{\zeta}:\;
-\frac{ih}{4}(\zeta-\frac{1}{\zeta})\Phi_1(\zeta)-\frac{h}{4}(\zeta+\frac{1}{\zeta})
\Phi_3(\frac{1}{\zeta})=F(\frac{1}{\zeta})
\eeq
\beq\label{mh14}
\zeta\,\rightarrow \,-{\zeta}:\;
\frac{ih}{4}(\zeta-\frac{1}{\zeta}){\Phi_1(-{\zeta})-
\frac{h}{4}(\zeta+\frac{1}{\zeta})
\Phi_3(\frac{1}{\zeta})={ -F(-\zeta})}
\eeq
\begin{figure}[ht]
\begin{center}
\resizebox{4in}{!}{ 
\includegraphics{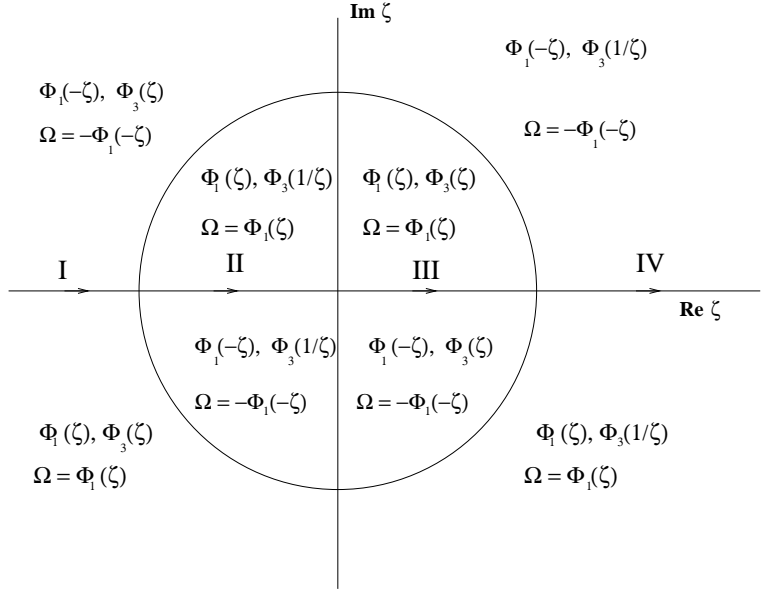}
}
\caption{Regions of analyticity of $\Phi$ and $\Omega$  
for the second Riemann-Hilbert problem for H equation}
\label{f10}
\end{center}
\end{figure}
The distribution of the relevant functions
in the $\zeta$-plane  is given in Figure {\ref{f10}}.
Introducing a new function $\Omega$ in terms of $\Phi_1$
as it is shown in Figure {\ref{f10}} we see that it has jumps on the real line and on the circumference. Using
(\ref{mh7}-\ref{mh14}) we can find its jumps on the real line
in terms of the known $F(\zeta)$ function on the intervals I-IV as follows
$$ I :\quad \Omega_+-\Omega_-=-\Phi_1(-\zeta)-
\Phi_1(\zeta)
=-\frac{4\zeta}{ih(\zeta^2-1)}(F(-\frac{1}{\zeta})+F(\zeta))$$
\beq\label{mh15}
II :\quad \Omega_+-\Omega_-=\Phi_1(-\zeta)+\Phi_1(\zeta)
=-\frac{4\zeta}{ih(\zeta^2-1)}( F(-\zeta)+F(\frac{1}{\zeta}))
\eeq
$$III:\quad \Omega_+-\Omega_-=\Phi_1(\zeta)+\Phi_1(-\zeta)=\frac{4\zeta}{ih(\zeta^2-1)}(F(\zeta)+ F(-\frac{1}{\zeta}))$$
$$IV:\quad \Omega_+-\Omega_-=-\Phi_1(\zeta)-
\Phi_1(-\zeta)=\frac{4\zeta}{ih(\zeta^2-1)}(F(\frac{1}{\zeta})+ F(-\zeta))$$
However we cannot do the same on  the circumference.
In a similar way introducing a new function ${\tilde\Omega}$ in terms of  $\Phi_3$  we can find jump functions on the imaginary axis, but not on the circle. Therefore
 to set up Riemann-Hilbert problem  for any of these functions we have to supplement the boundary
conditions with some physical,  for example Sommerfeld's radiation
conditions \cite{g6}.  That means that we have to 
estimate the asymptotic value of $u$ ( {\ref{h421a}})  for big $R$, where $R^2=x^2+z^2$
($x=R\cos \theta,\;z=R\sin\theta$) we will use the steepest descent method.
Let's introduce 
\beq\label{h421b}
E(\zeta )= \frac{ih}{2}(\zeta 
+\frac{1}{\zeta})z-
\frac{ h}{2}(\zeta-\frac{1}{\zeta})x
\eeq
Switching to $R$ it can be rewritten as
\beq\label{h421c}
E(\zeta )= \frac{iRh}{2}((\zeta+\frac{1}{\zeta})\sin \theta+i(\zeta-\frac{1}{\zeta})\cos\theta)
\eeq
Then from the equation $E'(\zeta)=0$ one obtains two stationary phase
points
 \beq\label{h421d}
\zeta_1= \sin\theta -i\cos \theta,\; \zeta_2=-\zeta_1
\eeq
Taking into account that we are considering a quarter-space $x,z\geq 0$, which means $0\leq \theta\leq \pi /2$, one can see that $\zeta_1$ is located on
$C_4$ and $\zeta_2$ is located on $C_2$.  It means that only these two integrals in (\ref{h421a})  will contribute in
the asymptotic value of $u$.  One can easily obtain that
 \beq\label{h421e}
E(\zeta_1)= iRh,\; E''(\zeta_1)=\frac{iRh}{(\sin \theta - i \cos \theta )^2}=
-iRhe^{-2i\theta}
\eeq
This in turn implies that in the neighborhood of the point $\zeta_{1}$ the
exponent $E(\zeta)$ takes the form,
\beq\label{h421f}
E(\zeta) \sim E(\zeta_1) +\frac{1}{2}E''(\zeta_1)(\zeta -\zeta_{1})^2
= iRh -\frac{1}{2}iRhe^{-2i\theta}(\zeta -\zeta_{1})^2,
\eeq
where $\zeta$ lies on the line tangent to the arc $C_{4}$ at the point
$\zeta_{1}$. Accordingly, the integral over $C_{4}$ can be estimated as
\beq\label{h421g1}
I_{C_4}=\frac{1}{2\pi i}\int_{C_4} \frac{e^{\frac{ih}{2}(\zeta 
+\frac{1}{\zeta})z-
\frac{ h}{2}(\zeta-\frac{1}{\zeta})x}\rho_{31}({\zeta})d \zeta}{\zeta}
\sim 
\frac{1}{2\pi i}\frac{\rho_{31}(\zeta_{1})}{\zeta_{1}}\int_{e^{i\theta}\real} e^{iRh 
-\frac{1}{2}iRhe^{-2i\theta}(\zeta-\zeta_1)^2}d \zeta
\eeq
Changing variables as $\zeta' = e^{-i\theta}(\zeta-\zeta_1),\; d\zeta'= e^{-i\theta}d\zeta $ and using again $\zeta$ for $\zeta '$ one obtains
\beq\label{h421h}
I_{C_4}= \frac{1}{2\pi i}\frac{\rho_{31}(\zeta_{1})}{\zeta_{1}} e^{iRh}e^{i\theta}\int_{\real}
e^{-\frac{1}{2}iRh\zeta^2}d \zeta
\eeq
Finally introducing $X=\sqrt{\frac{Rh}{2}}\zeta$ we can finish the estimate as
\beq\label{h421i}
I_{C_4}= \frac{1}{2\pi i}\frac{\rho_{31}(\zeta_{1})}{\zeta_{1}} e^{iRh}e^{i\theta}\sqrt{\frac{2}{Rh}}\int_{\real}e^{-iX^2}dX= \frac{1}{2\pi i}\frac{\rho_{31}(\zeta_{1})}{\zeta_{1}} e^{iRh}e^{i\theta}\sqrt{\frac{2}{Rh}}e^{-i\pi/4}\sqrt{\pi}
\eeq

Similarly, 
 \beq\label{h422e}
E(\zeta_2)=- iRh,\; E''(\zeta_1)=-\frac{iRh}{(-\sin \theta + i \cos \theta )^2}=
iRhe^{-2i\theta},
\eeq
and the integral over $C_{2}$ satisfies the asymptotic
relation 
\beq\label{h422i}
I_{C_2}= \frac{1}{2\pi i}\frac{\rho_{13}(\zeta_{1})}{\zeta_{1}} e^{-iRh}e^{i\theta}\sqrt{\frac{2}{Rh}}\int_{\real}e^{iX^2}dX= \frac{1}{2\pi i}\frac{\rho_{13}(\zeta_{1})}{\zeta_{1}} e^{-iRh}e^{i\theta}\sqrt{\frac{2}{Rh}}e^{i\pi/4}\sqrt{\pi}
\eeq

Now, taking into account the radiation condition, we arrive at the 
equation,
\beq\label{xxx}
\rho_{31}(\zeta) = 0, \quad \forall \zeta \in C_{2}\eeq
That means that (\ref{mh7}) holds in $C_2$:
\beq\label{mh111}
\frac{ih}{4}(\zeta-\frac{1}{\zeta})\Phi_1(\zeta)-\frac{h}{4}(\zeta+\frac{1}{\zeta})
\Phi_3(\zeta)=F(\zeta)
\eeq
We could think about this equation as about a jump between $\Phi_3$ which is analytic outside of $C_2$ and $\Phi_1$ which is analytic inside (see Figure{ \bf{2}}). However, our goal is to   write the jump for the function $\Omega$ on $C_2$
which we need  to  supplement the auxiliary Riemann-Problem for this function. One can see that   $\Phi_1(-\zeta)$ is also analytic outside
$C_2$ and is related to $\Phi_3$ by (\ref{mh12}). Therefore, this equation can be used   to express $\Phi_3$ in terms of  $\Phi_1(-\zeta)$:
\beq\label{mh121}
-\frac{h}{4}(\zeta+\frac{1}{\zeta})
\Phi_3(\zeta)=-F(-\frac{1}{\zeta})
+\frac{ih}{4}(\zeta-\frac{1}{\zeta})\Phi_1(-\zeta)
\eeq
Finally, substituting (\ref{mh121}) into (\ref{mh111})  we obtain the following jump for the  function $\Omega$ on $C_2$:
$$C_2 :\quad \Omega_+-\Omega_-=-\Phi_1(\zeta) - \Phi_1(-\zeta)
=-\frac{4\zeta}{ih(\zeta^2-1)}( F(\zeta)+F(-\frac{1}{\zeta})$$
Using the symmetry of $\Phi$ functions in the same way as we did before
we could obtain the similar relations for $C_3$ , $C_4$ and
$C_1$ and obtain the jump function $\Omega$ on the whole circle:
$$C_3,\,\, C_4 :\quad \Omega_+-\Omega_-
=\frac{4\zeta}{ih(\zeta^2-1)}( F(-\zeta)+F(\frac{1}{\zeta}))$$
$$C_1,\,\, C_2 :\quad \Omega_+-\Omega_- = -\frac{4\zeta}{ih(\zeta^2-1)}( F(\zeta)+F(-\frac{1}{\zeta})$$

Our  principal message now is the following: 

\vskip .2in
\noindent
{\it The global relations together with
the Sommerfeld radiation condition provide the complete set of the jump relation
for the unknown function $\Omega$}. 

\vskip .2in
\noindent
The function $\Omega(\zeta)$, can be now written in the form of the Cauchy
integral, 
 \begin{equation}\label{final}
 \Omega = \frac{1}{2\pi i}\int_\Gamma\frac{r(\zeta)}{\zeta'-\zeta}d\zeta,
 \end{equation}
 where the oriented  contour $\Gamma$ consists of the real line and the unit circle, and the 
 density function $r(\zeta)$ is given by the equations,
 $$
 r(\zeta) = -\frac{4\zeta}{ih(\zeta^2-1)}( F(\zeta)+F(-\frac{1}{\zeta}))
 $$
 on the parts $I$, $C_1$, $C_2$ and $-III$  of the contour $\Gamma$, and
$$
 r(\zeta) = \frac{4\zeta}{ih(\zeta^2-1)}( F(-\zeta)+F(\frac{1}{\zeta}))
 $$
 on the parts $IV$, $C_3$, $C_4$ and $-II$  of the contour $\Gamma$. 
 
 Equation (\ref{final})  completes the solution of the boundary value problem (\ref{mh1}). It would be
 very interesting  to compare the method of this paper with the alternative approach developed in 
 in \cite{spence} and \cite{ssf} for the quarter-plane problem for the same Helmholtz equation.

\vskip 2.in

\section*{Acknowledgment}

This work was partially supported by the National Science 
Foundation (NSF) under Grants No. DMS-0203104, No. DMS-0701768, No. DMS-1001777 and by a grant of the London Mathematical Society.

\end{document}